\documentclass[useAMS,usenatbib]{mnras}
 \usepackage{graphicx}
 \usepackage{epstopdf}
 \usepackage[colorinlistoftodos]{todonotes}
 \usepackage{natbib}
 \usepackage{morefloats}
 \usepackage{amsmath}
 \usepackage{dirtytalk}
\newcommand{\REAL}{{\rm I\!R}}

\title[21cm Emulation and Parameter Inference]{Emulation of reionization simulations for Bayesian inference of astrophysics parameters using neural networks}
\author[C.J.Schmit and J.R.Pritchard]{C. J. Schmit$^{1}$\thanks{E-mail:
claude.schmit13@imperial.ac.uk} and J.~R.~Pritchard$^{1}$\thanks{E-mail:
j.pritchard@imperial.ac.uk}\\
$^{1}$Blackett Laboratory, Imperial College, London, SW7 2AZ, UK}
\begin{document}



\maketitle

\label{firstpage}

\begin{abstract}

Next generation radio experiments such as LOFAR, HERA and SKA are expected to probe the Epoch of Reionization and claim a first direct detection of the cosmic 21cm signal within the next decade. 
Data volumes will be enormous and can thus potentially revolutionize our understanding of the early Universe and galaxy formation.  
However, numerical modelling of the Epoch of Reionization can be prohibitively expensive for Bayesian parameter inference and how to optimally extract information from incoming data is currently unclear. 
Emulation techniques for fast model evaluations have recently been proposed as a way to bypass costly simulations.
We consider the use of artificial neural networks as a blind emulation technique.
We study the impact of training duration and training set size on the quality of the network prediction and the resulting best fit values of a parameter search.
A direct comparison is drawn between our emulation technique and an equivalent analysis using 21CMMC.
We find good predictive capabilities of our network using training sets of as low as 100 model evaluations, which is within the capabilities of fully numerical radiative transfer codes.
\end{abstract}


\section{Introduction} 
\label{sec: intro}

Observations of the redshifted 21cm line of neutral hydrogen from the Epoch of Reionization (EoR) promise new insights into our understanding of the astrophysics of the first galaxies. 
Data is now becoming available from instruments such as LOFAR \citep{Patil2017}, MWA \citep{Dillon2015}, PAPER \citep{Ali2015}, and HERA \citep{DeBoer2017} such that increasingly stringent upper limits can be placed on reionization scenarios.
Data sensitivity and volume are bound to increase with SKA taking its first data in the early 2020s.
One of the challenges all these observations face is how to infer astrophysical parameters from 21cm observations. 
A common approach is to attempt Bayesian inference, typically implemented using MCMC techniques \citep{greig2015, greig2017, Harker2012, Hassan2017}. 
Such methods require an evaluation of the reionization model, typically a computationally expensive numerical simulation, many thousands of times.
This can be prohibitively expensive for the use of fully numeric simulations, and in order to make the inference tractable, one typically uses approximate semi-numerical simulations \citep{mesinger2007,Santos2010,mesinger2011,Fialkov2012}. 
These models sacrifice accuracy in favour of speed and thus limit the physics that can be included in the model.
Therefore, alternative techniques for speeding up the inference using fully numerical models are desirable. 
 
\citet{shimabukuro2017} recently proposed the use of artifical neural networks (ANN) for parameter estimation from 21cm observations. 
In that work, a training set of semi-numerical simulations was used to train an ANN so that from the shape of the 21cm power spectrum the network could predict the corresponding set of parameters. 
In this paper, we consider the opposite problem, i.e. to predict the power spectrum from a set of parameters. 
This technique promises to speed up the parameter inference significantly by needing to run a full model simulation only a small number of times for a training set that the ANN can train on and subsequently use the ANN prediction for the model evaluation.


MCMC techniques typically require evaluation of many closely spaced points in parameter space to fully sample from the posterior. 
This is computationally wasteful, since in many cases the simulation output varies smoothly and courser sampling would be sufficient to map the shape of the output, e.g. the 21cm power spectrum.
We therefore explore the possibility of emulating the output of the simulation. 
If we have a simulation $y = f(x)$, where $f$ is expensive to calculate, we can seek some approximate calculation $\tilde{y} = \tilde{f}(x)$, where $\tilde{f}$ is a fast emulation of the true simulation and the difference between $y$ and $\tilde{y}$ can be made as small as desired.
For our purposes, we seek to emulate the calculation of the 21cm power spectrum in a number of specified $k$ bins i.e. $y = \left\lbrace P(k_i|\boldsymbol{\theta})\right\rbrace$, where the subscript specifies the bin, for a restricted set of astrophysics parameters $\boldsymbol{\theta} = \left(\zeta, T_{\rm vir}, R_{\rm mfp}\right)$.
We can then use our emulation $\tilde{P}(k_i|\boldsymbol{\theta})$ to make rapid evaluations of the likelihood, 
\begin{equation}\label{eq:loglikelihood}
\ln \mathcal{L} = \sum\limits_i\frac{\left[P_{\rm obs}(k_i) - P(k_i|\boldsymbol{\theta})\right]^2}{2P_N(k_i)},
\end{equation}
where $P_{\rm obs}$ is an observed or mock data set, and $P_N$ is the noise power spectrum associated with a specified instrument.

Emulation techniques have been used in cosmology before. 
For example, \citet{heitmann2009, heitmann2014, heitmann2016} made use of Latin hypercube sampling (LHS) coupled to gaussian processes for regression to accurately emulate the numerical output of N-body simulations for the non-linear density power spectrum. 
Latin hypercube sampling techniques sample all parameters uniquely in all dimensions, this prevents wasteful model evaluations at already sampled parameter values. 
Alternatively, \citet{Agarwal2012} have made use of neural networks for the same purpose in their PKANN simulations.
As this paper was being completed, \citet{Kern2017} suggested emulation and the use of gaussian processes for the field of 21cm cosmology. 
In their paper they find a significant speed up for parameter searches while retaining a high degree of precision as compared to the brute force MCMC evaluation of \citet{greig2017}.

In this paper, we make use of neural networks to emulate the output of \texttt{21cmFast}\footnote{Publically available at \newline \texttt{https://github.com/andreimesinger/21cmFAST.}} and study the effect of the training set size on the predictive power of the emulator.
We aim to directly compare the performance of our emulator to the results of \citet{greig2015}, and therefore utilize the same $\Lambda$CDM parameter set. 
We fix the cosmology with ($\Omega_M$, $\Omega_\Lambda$, $\Omega_b$, $n_S$, $\sigma_8$, $H_0$) $=$ (0.27, 0.73, 0.046, 0.96, 0.82, 70 km s$^{-1}$ Mpc$^{-1}$).
An updated set of parameters, conforming with the latest \citet{PlanckCollaboration2015r_13} results could and should be used for future analyses. 
We introduce the general theory of neural networks in Section \ref{sec: neural} and our physical model in Section \ref{sec: model}. 
We then test our network and compare it to the model in Section \ref{sec: power}. 
Our main aim is a comparison of a Bayesian parameter search, which we introduce is Section \ref{sec: bayesian}, between our emulation technique and a brute-force MCMC search. 
We present our findings in Section \ref{sec: Discussion} and finally conclude in Section \ref{sec: conclusion}.

\section{Neural networks} 
\label{sec: neural}
In this section we give a general outline of the neural network used in our simulations (eg. \cite{Cheng1994}).
 
\subsection{Architecture}\label{sec: architecture}
In this work we use a multilayer perceptron (MLP) as our neural network design. 
MLPs use the supervised learning paradigm, where a set of training data $T \subset X \times Y$, where $X$ denotes the input or parameter space and $Y$ denotes the output space, is provided and upon which the neural network tries to fit a mapping $f: X\rightarrow Y$.
This is to say that the neural network is finding a mapping between input and output data, which is sensitive to the key features of the training set. 
This mapping can then be used on unknown data where the neural network uses its acquired knowledge of the system to infer an output, either in form of a classification or a number.

\begin{figure}
\includegraphics[width=0.5\textwidth]{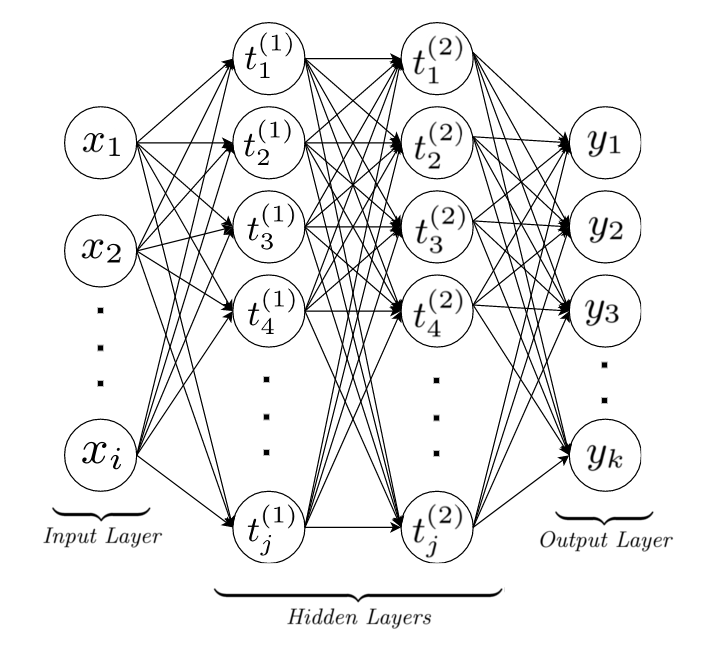}
\caption{Multilayer perceptron layout.}
\label{fig: Neural Network architecture}
\end{figure}

A neural network consists of three types of layers each consisting of a set of nodes or neurons, illustrated in Figure \ref{fig: Neural Network architecture}.   
The input layer takes $N_i$ data points into $N_i$ input nodes from which we want to predict some output.
Each node in the input layer is connected to all of $N_j$ nodes in the first of $L$ hidden layers via some weight $w^{(1)}_{ij}$.
The input to the nodes in the hidden layer is a linear combination of the input data and the weights,
\begin{equation}
s^{(1)}_j = \sum\limits_{i = 1}^{N_i} x_i w^{(1)}_{ij}.
\end{equation}
A neuron is then activated by some activation function $g: \REAL \rightarrow \REAL$. 
We use a sigmoid activation function, $g(s) = 1/(1+e^{-s})$, as this non-linear function allows us to fit to any function in principle \citep{Cybenko1989}. 
This activation step can be interpreted as each neuron having specialised on a certain feature in the system \citep{Bishop2006,Gal2016Uncertainty} and when the data reflects this feature the neuron will be activated.
The output from the neuron activation is then fed into the next hidden layer as input, such that the $j^\text{th}$ neuron in the $l^\text{th}$ hidden layer  computes,
\begin{equation}\label{eq: definition t}
t^{(l)}_j = g\left(s^{(l)}_j\right),
\end{equation}
where, for $1<l\leq L$,
\begin{equation}\label{eq: definition s}
s^{(l)}_j = \sum\limits_{i = 1}^{N_j}t^{(l-1)}_i w^{(l)}_{ij}.
\end{equation}
Finally, the output layer combines the outputs from the final hidden layer into $N_k$ desired output values,
\begin{equation}
y_k = \sum\limits_{i = 1}^{N_j}t^{(L)}_i w^{(L+1)}_{ik}.
\end{equation}

The weights between neurons $w^{(l)}_{ij}$ are obtained during the training of the network, where we apply the Limited-memory Broyden-Fletcher-Goldfarb-Shanno (LBFGS) algorithm \citep{Press2007} to minimize the mean-square error between the true value provided by the training data and the value predicted by the network. 
This training algorithm is ideal for sparse training sets and a low dimensional parameter space \citep{Le2011}, and will be discussed in the following section.
\subsection{Supervised learning}

A popular training algorithm for machine learning problems is back-propagation via gradient descent \citep{Rumelhart1986,Cheng1994,Abu-Mostafa2012, shimabukuro2017}.
However, back-propagation requires the user to manually set a learning rate which must fall within a finite range, too small and each training iteration produces vanishingly small changes, too large and the training steps overshoot. 
An arbitrary learning rate does therefore not guarantee that the network will converge to a point with vanishing gradient and second order optimization methods can be used to guarantee convergence \citep{Battiti1992a}.

Suppose we have $N_\text{train}$ training sets consisting of $N_i$ input parameters and $N_\text{out}$ output data. These training sets are fed into a neural network as described in the previous section. 
Training an ANN can then be viewed as an optimization problem where one seeks to minimize the total cost function $E(\boldsymbol{w})$, which is the sum-squared error over the training sets.
\begin{equation}\label{eq: definition E}
E(\boldsymbol{w}) = \sum\limits_{n = 1}^{N_\text{train}} E_n(\boldsymbol{w}) =\sum\limits_{n = 1}^{N_\text{train}} \left[\frac{1}{2} \sum\limits_{i = 1}^{N_\text{out}} \left(y_{i,n}(\boldsymbol{w}) - d_{i,n}\right)^2\right],
\end{equation}
where $y_{i,n}$ is the prediction made by the neural network in the $i^\text{th}$ neuron of the output layer, using the $n^\text{th}$ parameter set of all training inputs, $d_{i,n}$ is the true result for the $i^\text{th}$ neuron in the output layer corresponding to the $n^\text{th}$ parameter set, and thus $E_n$ is the cost function associated with the $n^\text{th}$ input  parameter set.

We can expand the cost function around some particular set of weights $\boldsymbol{w}_0$ using a Taylor series,
\begin{equation}
\begin{aligned}
E(\boldsymbol{w}) = & E(\boldsymbol{w}_0) + (\boldsymbol{w} - \boldsymbol{w}_0)^T \boldsymbol{g}_0\\
&+\frac{1}{2} (\boldsymbol{w} - \boldsymbol{w}_0)^T H_0 (\boldsymbol{w}-\boldsymbol{w}_0) +  ... ,
\end{aligned}
\end{equation}
where $\boldsymbol{g}_0$ is the vector of gradients and $H_0$ denotes the Hessian matrix with elements, 
\begin{equation}
h_{ij} = \frac{\partial^2 E}{\partial w_i \partial w_j}.
\end{equation} 
Whereas back-propagation is based on a linear approximation to the error surface, better performance can be expected when using a quadratic error model,
\begin{equation}
\begin{aligned}
E(\boldsymbol{w}) \approx &E(\boldsymbol{w}_0) + (\boldsymbol{w} - \boldsymbol{w}_0)^T \boldsymbol{g}_0\\
&+ \frac{1}{2}(\boldsymbol{w} - \boldsymbol{w}_0)^T H_0 (\boldsymbol{w}-\boldsymbol{w}_0).
\end{aligned}
\end{equation}
Provided $H_0$ is positive definite, this approximation to the error surface has a minimum, $\partial E/\partial \boldsymbol{w} = 0$, at
\begin{equation}
\boldsymbol{w} = \boldsymbol{w}_0 - H_0^{-1}\boldsymbol{g}_0.
\end{equation} 
Given that a quadratic approximation to the actual cost function is used, an iterative approach needs to be taken in order to find an estimate of the true minimum.
Similar to back-propagation where $\boldsymbol{g}$ is used as the search direction, second order methods use $ -H^{-1}\boldsymbol{g}$ as the search direction.
Thus the search direction during training iteration $k$ is given by,
\begin{equation}
\boldsymbol{\Delta}_k = - H_k^{-1}\boldsymbol{g}_k.
\end{equation}
Solving this system of equations requires precise knowledge of the Hessian, as well as a well-conditioned Hessian, which in is not always guaranteed. 
Instead of computing the Hessian and inverting it, the BFGS scheme seeks to estimate $H^{-1}_k$ directly from the previous iteration.
\cite{McLoone2002} give the basic algorithmic structure as follows;
\begin{itemize}
\item Set the search direction $\boldsymbol{\Delta}_{k-1}$ equal to $-M_{k-1} \boldsymbol{g}_{k-1}$, where $M_{k-1}$ is the approximation to $H^{-1}_{k-1}$ at the ($k-1$)$^\text{th}$ iteration.
\item Use a line search to find the weights which yield the minimum error along $\boldsymbol{\Delta}_{k-1}$,
\begin{equation}
\boldsymbol{w}_k = \boldsymbol{w}_{k-1} + \eta_{opt} \boldsymbol{\Delta}_{k-1},
\end{equation}
\begin{equation}
\eta_{opt} = \min\limits_\eta(E(\boldsymbol{w}_{k-1} + \eta \boldsymbol{\Delta}_{k-1})).
\end{equation}
\item Compute the new gradient $\boldsymbol{g}_k$.
\item Update the approximation to $M_k$ using the new weights and gradient information.
\begin{equation}
\boldsymbol{s}_{k} = \boldsymbol{w}_{k}- \boldsymbol{w}_{k-1} \text{ and } \boldsymbol{t}_{k} = \boldsymbol{g}_{k} - \boldsymbol{g}_{k-1},
\end{equation}
\begin{equation}
A_k = \left(1 + \frac{\boldsymbol{t}^T_k M_{k-1}\boldsymbol{t}_{k}}{\boldsymbol{s}^T_k\boldsymbol{t}_k}\right)\frac{\boldsymbol{s}_k \boldsymbol{s}^T_k}{\boldsymbol{s}^T_k\boldsymbol{t}_k},
\end{equation}
\begin{equation}
B_k = \frac{\boldsymbol{s}_k\boldsymbol{t}^T_k M_{k-1} + M_{k-1}\boldsymbol{t}_k\boldsymbol{s}_k}{\boldsymbol{s}^T_k\boldsymbol{t}_k},
\end{equation}
\begin{equation}\label{eq: update}
M_k = M_{k-1} + A_k - B_k.
\end{equation}
\end{itemize}
The scheme initializes by taking a step in the direction of steepest descent by setting, $M_0 = \mathcal{I}$.

The limited-memory BFGS scheme we are using, recognizes the memory intensity of storing large matrix estimates of the inverse Hessian, and resets $M_{k-1}$ to the identity matrix in equation \eqref{eq: update} at each iteration and multiplies through by $-\boldsymbol{g}_k$ to obtain a matrix free expression for $\boldsymbol{\Delta}_k$. 
\begin{itemize}
\item The LBFGS thus uses the following update formula \citep{Asirvadam2004},  
\begin{equation}
\boldsymbol{\Delta}_k = - \boldsymbol{g}_k + a_k \boldsymbol{s}_k + b_k \boldsymbol{t}_k,
\end{equation}
with, 
\begin{equation}
a_k = -\left(1 + \frac{\boldsymbol{t}^T_k \boldsymbol{t}_{k}}{\boldsymbol{s}^T_k\boldsymbol{t}_k}\right)b_k + \frac{\boldsymbol{t}^T_k \boldsymbol{g}_k}{\boldsymbol{s}^T_k\boldsymbol{t}_k} \text{ and } b_k = \frac{\boldsymbol{s}^T_k \boldsymbol{g}_k}{\boldsymbol{s}^T_k\boldsymbol{t}_k}.
\end{equation}
\end{itemize}

\section{Reionisation Model} 
\label{sec: model}

In order to produce the training sets upon which our neural network is ultimately trained, we need to model the EoR and the 21cm power spectrum as a function of some tangible model parameters. 

The main observable of 21cm studies is the 21cm brightness temperature, defined by \citep{Pritchard2012, Furlanetto2006}, 
\begin{equation}
\begin{aligned}
\delta T_b (\nu) &= \frac{T_S - T_\gamma}{1+z} (1- e^{-\tau_{\nu_0}}) \\
&\approx 27 x_\text{HI} (1+\delta_b)\left(\frac{\Omega_b h^2}{0.023}\right) \left(\frac{0.15}{\Omega_M h^2}\frac{1+z}{10}\right)^{1/2}\\
&\times \left(1 - \frac{T_\gamma(z)}{T_S}\right) \left[\frac{\partial_r v_r}{(1+z) H(z)}\right]^{-1} \text{mK},
\end{aligned}
\end{equation}
where $x_\text{HI}$ denotes the neutral fraction of hydrogen, $\delta_b$ is the fractional overdensity of baryons, $\Omega_b$ and $\Omega_M$ are the baryon and total matter density in units of the critical density, $H(z)$ is the Hubble parameter and $T_\gamma(z)$ is the CMB temperature at redshift $z$, $T_S$ is the spin temperature of neutral hydrogen, and $\partial_r v_r$ is the velocity gradient along the line of sight. 
One can define the 21cm power spectrum from the fluctuations in the brightness temperature relative to the mean, 
\begin{equation}
\delta_{21}(\boldsymbol{x}, z) \equiv \frac{\delta T_b(\boldsymbol{x}) - \left\langle\delta T_b\right\rangle}{\left\langle\delta T_b\right\rangle},
\end{equation}
where $\langle...\rangle$ takes the ensemble average.
The dimensionless 21cm power spectrum, $\Delta^2_{21}(k)$, is then defined as, 
\begin{equation}
\Delta^2_{21}(k) =  \frac{k^3}{2\pi^2}P_{21}(k),
\end{equation}
where $P_{21}(k)$ is given through, 
\begin{equation}
\left\langle\tilde{\delta}_{21}(\boldsymbol{k})\tilde{\delta}_{21}(\boldsymbol{k'})\right\rangle = (2\pi)^3 \delta^D(\boldsymbol{k}-\boldsymbol{k'})P_{21}(k).
\end{equation}
Here, $\tilde{\delta}_{21}(\boldsymbol{k})$ denotes the Fourier transform of the fluctuations in the signal and $\delta^D$ denotes the 3D Dirac delta function.

The 21cm power spectrum is the most promising observable for a first detection of the signal \citep{Furlanetto2006}, and encodes information about the state of reionization throughout cosmic history. 
For the evaluation of the 21cm power spectrum we utilize the streamlined version of \texttt{21cmFast}, which was used in the MCMC parameter study of \citet{greig2015}.
This version of \texttt{21cmFast} is optimized for astrophysical parameter searches.

The astrophysical parameters that we allow to vary in our model are three-fold. 

\textit{Ionizing efficiency}, $\zeta$: The ionization efficiency combines a number of reionization parameters into one.
We define $\zeta = A_\text{He}f_* f_\text{esc} N_\text{ion}$, where $A_\text{He} = 1.22$ is a correction factor to account for the presence of helium and converts the number of ionizing photons to the number of ionized hydrogen atoms, $f_*$ is the star formation efficiency, $f_\text{esc}$ is the escape fraction for UV radiation to escape the host galaxy, and $N_\text{ion}$ is the number of ionizing photons per baryons produced. 
These parameters are poorly constrained at high redshifts.
As $N_\text{ion}$ depends on the metallicity and the initial mass function of the stellar population, we can approximate $N_\text{ion} \approx 4000$ for Population II stars with present day initial mass function, and $N_\text{ion} < 10^4$ for Population III stars. 
The value for the star formation efficiency $f_*$ at high redshifts is extremely uncertained due to the lack of collapsed gas. 
Therefore, although $f_* \approx 0.1$ is reasonable for the local Universe it is uncertain how this relates to the value at high redshifts. 
Additionally a constant star formation rate has been disfavoured by recent studies \citep{Mason2015,Mashian2015a,Furlanetto2017}.
For our purpose however, a simplistic constant star formation model is sufficient.
Similarly, the UV escape fraction $f_\text{esc}$ observed for local galaxies only provides a loose constraint for the high redshift value. 
Although $f_\text{esc} < 0.05$ is reasonable for local galaxies, large variations within the local galaxy population is observed for this parameter.
We thus allow the ionization efficiency to vary significantly in our model to reflect the uncertainty on the limits of this parameter, and consider $5 \leq \zeta \leq 100$. 

\textit{Maximal distance travelled by ionizing photons}, $R_\text{mfp}$: As structure formation progresses, dense pockets of neutral hydrogen gas emerge where the recombination rate for ionized proton - electron pairs is much higher than the average IGM. 
These regions of dense hydrogen gas are called Lyman limit systems (LLS) and effectively absorb all ionizing radiation at high redshifts. 
This effectively limits the bubble size of ionized bubbles during reionization.
EoR models include the effect of these absorption systems as a mean free path of the ionizing photons. 
However, due to the limited resolution of 21cmFAST, this sub-grid physics is modelled as a hard cut-off for the distance travelled by ionizing photons. 
As our allowed range for this parameter we use, $2 \text{ Mpc} \leq R_\text{mfp} \leq 20 \text{ Mpc}$.

\textit{Minimum virial temperature for halos to produce ionizing radiation}, $T_\text{vir}$:
Star formation is ultimately regulated by balancing thermal pressure and gravitational infall of gas in virialized halos. 
Molecular hydrogen allows gas to cool rapidly, on timescales lower than the dynamical timescale of the system, such that an unbalance of the two opposing forces occurs and the gas collapses which triggers a star to form. 
Although initial bursts of population III stars are thought to be able to occur briefly in halos virialized at $T_\text{vir} \sim 10^3\text{ K}$, these stars produce a strong Lyman-Werner background which leads to a higher dissociation of H$_2$ molecules. 
Star formation then moves to halos with $T_\text{vir} > 10^4\text{ K}$, where HI is ionized by virial shocks and atomic cooling is efficient.
$T_\text{vir}$ thus sets the threshold for star formation and we consider $10^4\text{ K} < T_\text{vir} < 2\times10^5\text{ K}$.
\section{Predicting the 21cm power spectrum} 
\label{sec: power}


We use two different approaches to emulate the 21cm power spectrum. 
First, we use a simple two-layer MLP, as described in Section \ref{sec: neural}, with 30 nodes in each layer, as we require the network to be sufficiently complex to map our set of 3 parameters to 21 power spectrum k-bins. 
This NN is then trained on a variety of training sets, see \ref{sec: grid training} and \ref{sec: LHC}, obtained from \texttt{21cmFast} simulations. 
Then, for comparison we use trilinear interpolation of the training set, simply interpolating the power spectrum on a parameter grid.

\subsection{Grid-based approch}\label{sec: grid training}
In order to study the impact of the choice of training set on the predicting power of the ANN we prepared a variety of training sets. 
The most basic approach is to distribute parameter values regularly in parameter space and obtaining the power spectrum for each point on a grid. 
We vary our parameters as per Section \ref{sec: model}, $5 \leq \zeta \leq 100 $, $10^4\text{ K} \leq T_\text{vir} \leq 10^5 K$ and $2$ Mpc $\leq R_\text{mfp} \leq 20$ Mpc, as these reflect our prior on the likely parameter ranges (see Section \ref{sec: model}).
Each training set then consists of the power spectrum evaluated in 21 $k$-bins, set by the box size of 250 Mpc, upon which the ANN is trained. 
We compare 5 different training sets at 2 different redshift bins, $z = 8$ and $z = 9$. 
These training sets consist of 3, 5, 10, 15 and 30 points per parameter, which leads to training sets of total size 27, 125, 1000, 3375 and 27000 respectively.

This approach is the most basic and certainly the most straight forward to implement, however it comes with a number of drawbacks. 
Projected down, a gridded set of parameter values has multiple points which occupy the same parameter values.  
This implies that the simulation is evaluated multiple times at the same values for some parameter at each point in any given row in the grid, see Figure \ref{fig: sampling}. 
Furthermore, if the observable is varying slowly in some parameter, few points are needed to model its behaviour and thus valuable simulation time is wasted on producing points in the grid that add very little information.

Another important limitation is the exponential scaling of the total number of points with the number of parameters in the grid.
In the simple three dimensional case which we are studying here, $N$ evaluations per parameter lead to a total of $N^3$ points on the grid. 
Ultimately, it is desirable to allow the model cosmology to vary and include at least 6 cosmological parameters into the search as well as additional astrophysical parameters, such as the X-ray efficiency, $f_X$, obscuring frequency, $\nu_\text{min}$, and the X-ray spectral slope, $\alpha_X$. 
One is then looking at a total of 12 or more parameter dimensions for which evaluations on the grid are prohibitively expensive and other techniques are needed.
A further problem is presented by the proportion of volume in the corners of a hyper-cubic parameter space\footnote{In 12 dimensions the proportion of the volume in the corners of a hypercube is $\sim 99.96\%$. 
That is the difference between the volume of the hypercube and that of an n-ball.}. 
High dimensional parameter spaces thus profit greatly by using hyperspherical priors which decrease the number of model evaluations in the low likelihood corner regions of parameter space drastically.

\subsection{Latin Hypercube approach}\label{sec: LHC}
\begin{figure}
\includegraphics[width=0.5\textwidth]{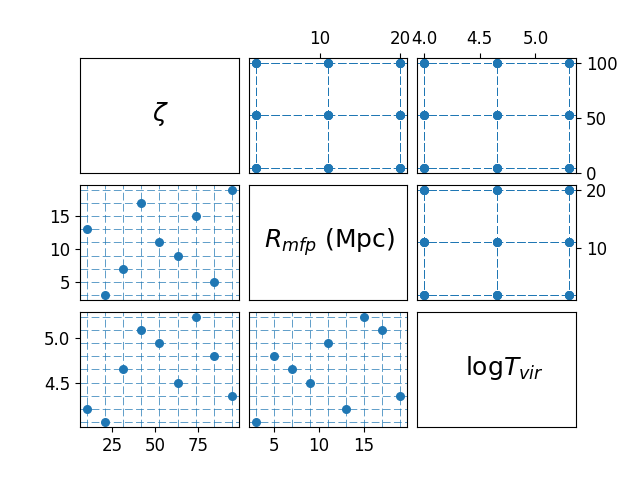}
\caption{Visualization of the two training techniques. The parameter space is projected down to two dimensions in each plot. Top right: 27 regularly gridded parameters. Bottom left: 9 samples which are obtained using the latin hypercube sampling technique.
Note that the number of samples are chosen such that the same number of projected samples are visible.}
\label{fig: sampling}
\end{figure}

A second approch is to use the latin hypercube sampling (LHS) technique, shown in Figure \ref{fig: sampling}. 
Here, the parameter space is divided more finely, such that no two assigned samples share any parameter value.
In two dimensions this method is equivalent to filling a chess board with rooks in such a way that no two of them threaten each other. 
Immediately, one of the shortcomings of the gridded parameter space is dealt with, in that the simulation need never be run at the same parameter value twice. 
The other main advantage of the LH is that its size does not increase exponentially with the dimension of parameter space.
This property makes the LH the only feasible way of exploring high dimensional parameter spaces with ANNs \citep{Urban2010}. 

We use a maximin distance design for our latin hypercube samples \citep{Morris1995a}. 
These designs try to simultaneously maximize the distance between all site pairs while minimizing the number of pairs which are separated by the same distance \citep{Moore1990a}.
This maximin design for LHS prevents highly clustered sample regions and ensures homogeneous sampling.
Prior knowledge of the behaviour of the power spectrum could also be used to identify the regions of parameter space where the power spectrum varies most rapidly and thus a higher concentration  of samples should be imposed on such a region.
Additionally, using a spherical prior region may help reducing the number of model evaluations used in the corners of parameter space where the likelihood is low \citep{Kern2017}.

For our training set comparisons we use 3 different LH training sets of size 100, 1000 and 10000 respectively.

\subsection{Power Spectrum Predictions}

\begin{figure}
\includegraphics[width=0.5\textwidth]{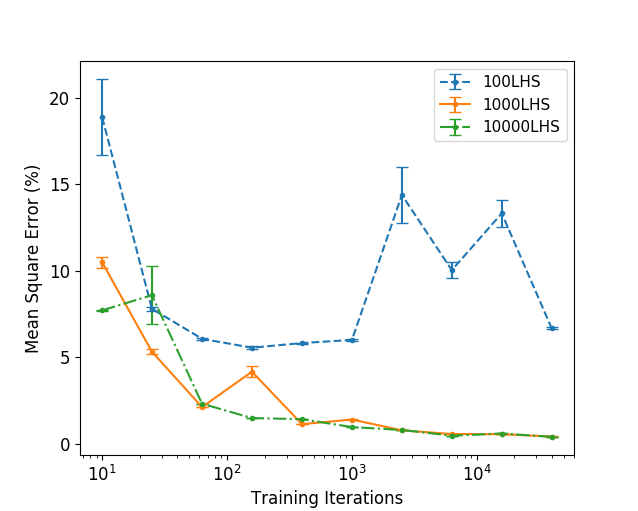}
\caption{Mean square error of the neural network prediction compared to a fixed test set of 50 points at $z = 9$ as a function of the training iterations.
At each number of training iterations, the training is repeated 10 times and we show the mean value of each resulting MSE and the variance on the mean as error bars. 
Shown are the behaviours for neural networks using 100 (blue), 1000 (orange) and 10000 (green) latin hypercube samples in the training set.}
\label{fig: training iter vs loss}
\end{figure}

We now test the predictive power of our trained ANN.
First, we define the mean square error between the true value of the power spectrum and an estimate given by the ANN,
\begin{equation}
\text{MSE} = \frac{1}{N_p N_k} \sum\limits_{i = 1}^{N_p}\sum\limits_{j = 1}^{N_k} \left(\frac{P^\text{true}_i(k_j) - P^\text{estimate}_i(k_j)}{P^\text{true}_i(k_j)}\right)^2,
\end{equation}
where $N_p$ is the number of parameter combinations we estimate and compare, and $N_k$ the number of $k$-bins used in the comparison.
We produced a test set of 50 \texttt{21cmFast} power spectra at $z = 9$, sampled from a LH design to ensure a homogeneous spread in parameter space. 
This test set was then compared to a prediction from our ANN trained on three sizes of training sets, using 100, 1000 and 10000 samples distributed again using a LH design.
We vary the training duration on each set and compare the predictions to the true values of the test set in Figure \ref{fig: training iter vs loss}.
The error bars are obtained by selecting $75\%$ of the total points in the training set at random for the network regression. 
The network is then trained on this subset and a value for the MSE is obtained.
A new training sample is then selected at random and the process is repeated 10 times. 
The error bars thus signify the expected error from any given latin hypercube sampled training set of comparable size. 

In the case of $10^3$ and $10^4$ samples in the training set, the neural network quickly approaches a relative mean square error of less than $1\%$.
With more than $10^3$ training iterations, both training sets show a clear reduction in the training efficiency.
The 100 LHS curve is dominated at high training iterations by outlier parameter points which are particularly poorly constrained. 
We find that these outliers can affect the MSE heavily while having a relatively small effect on the final parameter inference. 
We define an outlier to be any $k$-bin whose square error is larger than 1, meaning a relative error of over 100\%. 
For a training set of 100 points, one should then expect up to $\sim 2\%$ of all $k$-bins to be outliers at any given training iteration.
This unexpected behaviour may indicate an insufficient coverage of the training set, or that our neural network retains a high degree of flexibility even after regressing over 100 training samples. 
For our training set of 1000 points, the fraction of outliers produced reduces to less than $\sim 1\%$, when the training iterations are low, and we cease to find any outliers at more than 100 training iterations.
This indicates a significant reduction in the freedom of the neural network and an increase of the confidence in our prediction. 
Of note is that some outliers have a greater impact than others and we find some whose square error $\sim 10$, indicating a complete failure to predict the power in that particular $k$-bin.
One should thus be cautious when using small training sets that may not sufficiently constrain the freedom of the neural network.
Based on the results for our two larger training sets, we proceed by using $10^4$ training iterations in all our neural network training.

\begin{figure}
\includegraphics[width=0.5\textwidth]{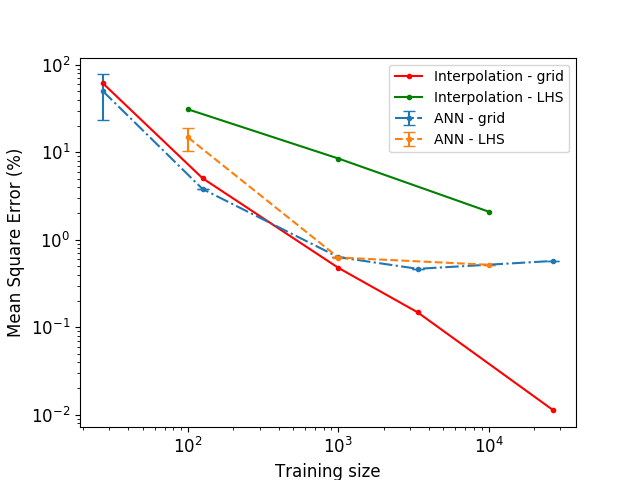}
\caption{Comparison between the mean square error of interpolation on a grid (red solid line), the neural network using gridded training sets (blue dot-dashed line), interpolation (green solid line) and the neural network using LHC training sets (orange dashed line). Neural networks are trained using $10^4$ training iterations. Plotted are the mean values after the NN is retrained 10 times, and the standard deviation to the mean is shown as error bars.}
\label{fig: comparison of errors}
\end{figure}

Further, we compare the mean square error between our training techniques against the training set size and sampling technique. 
In Figure \ref{fig: comparison of errors}, we compare the mean square error in the prediction when the gridded parameter values are interpolated (red), or used to train our neural network (blue), with the predictions obtained when using a Latin hypercube sampled training set (green and orange).
Similar to Figure \ref{fig: training iter vs loss}, we compute the mean and variance of the MSE over 10 separately trained networks by selecting $75\%$ of the samples in the training set at random at a time. 

As expected, when using a finer grid of parameters to interpolate the power spectrum, the accuracy of the prediction increases. 
Although the neural network predictions increase in accuracy for both the grid and the LHC, a clear plateauing in the addition of information by a larger training set can be observed.
We thus observe a fundamental limit to the relative mean square error for the neural network design. 
This limit depends on the design parameters of the neural network and can be optimized via $k$-fold validation of the networks design or hyper-parameters.
Varying the design parameters, such as the number of hidden layers or number of nodes per layer, and minimizing the mean square error for a power spectrum prediction over $k$ iterations can reduce networks error bound.
Our network design limits errors at $\sim 1\%$, which is sufficiently below any confidence limit associated with our model, that optimizing design parameters is of limited use.
Optimisation via $k$-fold validation may be necessary when using fully numerical simulations which reflect a higher degree of physical accuracy than fast semi-numerical methods.
No clear difference of the MSE can be seen comparing the latin hypercube sampled training sets and those produced on the grid in 3 dimensions.
We expect a more significant discrepancy in higher dimensions of parameter space as discussed in Section \ref{sec: LHC}. 
As such it is instructive to compare the performance of the interpolation on the grid to that on the LH. 
The ANN manages to capture the information of the unstructured training data much better than simple interpolation does, whereas this is not necessarily the case for large gridded training sets.

\begin{figure}
\includegraphics[width=0.5\textwidth]{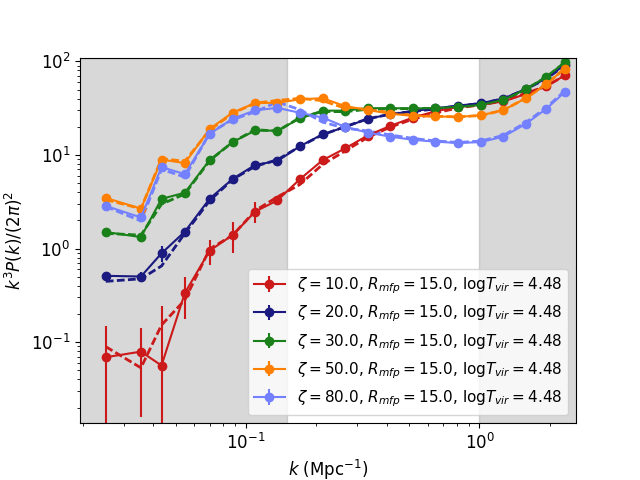}
\caption{Comparison between neural network prediction of the 21cm power spectrum (solid line) and the \texttt{21cmFast} power spectrum (dashed line). We vary $\zeta$ at $z = 9$ from $\zeta = 10$ to $\zeta = 80$, and use 1000 training iterations on 75\% of the 1000 LHS training set selected at random. This process is repeated 10 times and the mean values are shown with the variance on the mean as error bars.}
\label{fig: ANN pred 1}
\end{figure}

\begin{figure}
\includegraphics[width=0.5\textwidth]{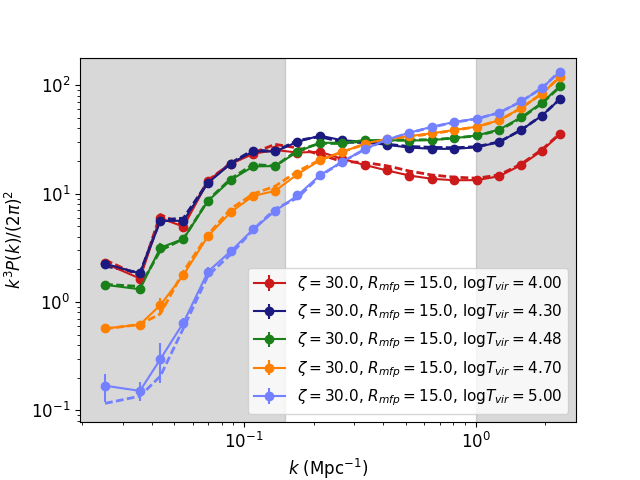}
\caption{Comparison between neural network prediction of the 21cm power spectrum (solid line) and the \texttt{21cmFast} power spectrum (dashed line). We vary $T_\text{vir}$ at $z = 9$ from $T_\text{vir}= 10^4$ K to $T_\text{vir} = 10^5$ K, similar to Figure \ref{fig: ANN pred 1}.}
\label{fig: ANN pred 2}
\end{figure}

\begin{figure}
\includegraphics[width=0.5\textwidth]{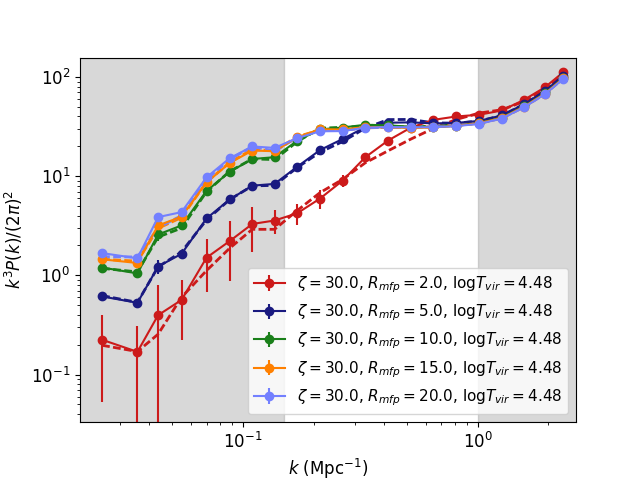}
\caption{Comparison between neural network prediction of the 21cm power spectrum (solid line) and the \texttt{21cmFast} power spectrum (dashed line). We vary $R_\text{mfp}$ at $z = 9$ from $R_\text{mfp}= 2$ Mpc to $R_\text{mfp} = 20$ Mpc, similar to Figure \ref{fig: ANN pred 1}.}
\label{fig: ANN pred 3}
\end{figure}

Figures \ref{fig: ANN pred 1} to \ref{fig: ANN pred 3} show the predictions of a trained neural network (solid lines) and the true values of the power spectrum at the same point in parameter space (dashed lines). 
In order to determine the dependence of the accuracy of the predictions on the particular training set used, a subset of the training set is again randomly selected and used as the training set.
Similar to before, the network is retrained 10 times while the predictions are averaged. 
The variance on the mean prediction in each $k$-bin is added as the expected error on the predicted mean value of the power spectrum.
The power spectrum is dominated on small scales ($k>1 \text{ Mpc}^{-1}$) by shot noise and by foregrounds on large scales ($k < 0.15 \text{ Mpc}^{-1}$). 
We therefore apply cuts at these scales in our analysis and indicate the noise dominated ranges by the grey shaded regions in figures \ref{fig: ANN pred 1}, \ref{fig: ANN pred 2} and \ref{fig: ANN pred 3}.

We observe that the network produces a good fit to the true values within the region of interest.
The size of the error bars indicates a very low dependence on the training subset used for training such that we conclude that the exact distribution of training sets in parameter space has little influence as long as it is homogeneously sampled.
We also observe that the network manages to fit $T_\text{vir}$ particularly well at large scales compared to the other two parameters whose error bars noticeably increase as $k$ approaches the foreground cut-off.
This shows that a sampling scheme that varies according to the dependence of the power spectrum on the input parameters may be advantageous to achieve some desired accuracy.

In the context of outliers, discussed earlier in this section, we see that the prediction for the power spectrum at $(\zeta, R_\text{mfp}, \log T_\text{vir}) = (30, 2, 4.48)$, in Figure \ref{fig: ANN pred 3}, overestimates the power at $k \approx 0.5 \text{ Mpc}^{-1}$ by a factor of $\sim 2$.
This point would have a relatively large impact on the MSE as recorded in Figure \ref{fig: training iter vs loss}, even though the network is very well behaved for most regions in parameter space. 


\section{Bayesian inference of astrophysical parameters.}
\label{sec: bayesian}
In Bayesian parameter inference one is interested in the posterior distribution of the parameters $\boldsymbol{\theta}$ within some model $\mathcal{M}$. 
That is the probability distribution of the parameters given some data set $\boldsymbol{x}$.
We can then write Bayes' Theorem,
\begin{equation}
Pr(\boldsymbol{\theta} | \boldsymbol{x}, \mathcal{M}) = \frac{Pr(\boldsymbol{x} | \boldsymbol{\theta}, \mathcal{M}) \pi(\boldsymbol{\theta}| \mathcal{M}) }{Pr(\boldsymbol{x}| \mathcal{M})},
\end{equation}
to relate the posterior distribution $Pr(\boldsymbol{\theta} | \boldsymbol{x}, \mathcal{M})$ to the Likelihood, $\mathcal{L} \equiv Pr(\boldsymbol{x} | \boldsymbol{\theta}, \mathcal{M})$, the prior, $\pi(\boldsymbol{\theta}| \mathcal{M})$, and a normalisation factor called the evidence, $Pr(\boldsymbol{x}| \mathcal{M})$. 
This expression parametrises the probability distribution of the model parameters as a function of the likelihood, which, given a model and a data set, can be readily evaluated under the assumption that the data points are independent and carry gaussian errors, 
\begin{equation}
\ln \mathcal{L} = -\frac{\left[\boldsymbol{x} - \boldsymbol{\mu}(\boldsymbol{\theta})\right]^2}{2\sigma_x^2} + C,
\end{equation}
where $C$ denotes a normalisation constant.
In our case, the data will be a mock observation of the 21cm power spectrum, $\boldsymbol{x} = \{P_\text{obs}(k_i)\}$, evaluated in 21 $k$-bins, the expectation value of the data will be the theoretical model prediction of the power spectrum, $\boldsymbol{\mu}(\boldsymbol{\theta}) = P(k, \boldsymbol{\theta})$, and for the variance on the data we assume that instrumental noise is the sole contributor characterised by a noise power spectrum, $\sigma_x^2 = P_\text{Noise}(k)$.

\subsection{Experimental Design}
We use \texttt{21cmSense}\footnote{Publically available at \texttt{https://github.com/jpober/21cmSense}.}  \citep{Pober2013,Pober2014} to compute the noise power spectrum for HERA331, with experimental details outlined in \citet{Beardsley2015} and summarized below.
The noise power spectrum used is given by \citep{Parsons2012},  
\begin{equation}
P_\text{Noise}(k) \approx X^2Y \frac{k^3}{2\pi^2}\frac{\Omega'}{2t} T_\text{sys},
\end{equation}
where $X^2Y$ denotes a conversion factor for transforming from the angles on the sky and frequency to comoving distance, $\Omega'$ is the ratio of the square of the solid angle of the primary beam and the solid angle of the square of the primary beam, $t$ is the integration time per mode, and $T_\text{sys}$ is the system temperature of the antenna, which is given by the receiver temperature of  100 K  plus the sky temperature $T_\text{sky} = 60 \left(\nu / 300 \text{ MHz}\right)^{-2.55}\text{ K}$.

As our experiment design, we assume a HERA design with 331 dishes distributed in a compact hexagonal array to maximize the number of redundant baselines, as HERA is optimized for 21cm power spectrum observations \citep{DeBoer2017,Liu2016}. 
Each dish has a diameter of 14 m, which translates into a total collecting area of $\sim 50950 \text{ m}^2$.
HERA antennas are not steered and thus use the rotation of the Earth to drift scan the sky.
An operation time of 6 hours per night is assumed for a total of 1000 hours of integration time per redshift. 
We consider both single redshift and multiple redshift observations assuming a bandwidth of 8 MHz. 
Although experiments like HERA and the SKA will cover large frequency ranges $\sim 50-250$MHz, foregrounds can limit the bandpass to narrower instantaneous bandwidths.

\subsection{MCMC}

We aim to compare our parameter estimation runs to those of \citet{greig2015} by using the same mock and noise power spectrum for HERA331 as input for our Neural Network parameter search.
Our fiducial parameter values are $\zeta = 30$, $R_\text{mfp} = 15 \text{ Mpc}$ and $T_{vir} = 30000\text{ K}$.

First, we perform an independent parameter search in two redshift bins, $z = 8$ and $z = 9$, the latter comparing directly to Figure 3 in \citet{greig2015}. 
The fiducial values for the average neutral fraction at these redshifts are $\bar{x}_\text{HI}(z = 8) = 0.48$ and $\bar{x}_\text{HI}(z = 9) = 0.71$.
For both the emulation and the 21CMMC runs we produce $2.1\times 10^5$ points in the MCMC chain for a like-for-like comparison between the two techniques.
 
Then, we analyse observations at redshifts $z = 8$,  $9$ and $10$ by combining the information in these redshift bins.
We take a linear combination of the $\chi^2$ statistics in each redshift bin.
Three separate ANNs are used for each redshift and are trained on the same training sets as for the individual redshift searches at $z = 8$ and 9.
The fiducial neutral fraction for our final mock observation is $\bar{x}_\text{HI}(z = 10) = 0.84$. 
A total of $2.1\times 10^5$ are again obtained both in the neural network search and the equivalent 21CMMC run. 

\section{Discussion} 
\label{sec: Discussion}

Similar to \citet{Kern2017}, we see a significant speed-up for the parameter estimation. 
For our fiducial chain size, we observe a speed up by 3 orders of magnitude for the sampling of the likelihood by emulation over the brute-force method.
Our 21CMMC runtime of 2.5 days on 6 cores for a single redshift is reduced to 4 minutes using the emulator.
In addition to the sampling, the Neural Network training requires on the order of $\sim 1$ minute for 100 training samples to $\sim 1$ hour for 10$^4$ training samples, which is not needed when evaluating the model at each point.
Compared to the total runtime of 21CMMC the training time presents a minor factor.

\subsection{Single Redshift Parameter Constraints}

\begin{table}
\begin{center}
\caption{Median values and 68\% confidence interval found in the parameter search via the brute-force method (21CMMC) and our ANN emulation at $z=9$ and $z = 8$. The fiducial parameter values for both redshifts are given by $(\zeta, R_\text{mfp}, \log T_\text{vir}) = (30,15,4.48)$.}
 \begin{tabular}{c @{\hskip 0.15cm} c @{\hskip 0.15cm} c @{\hskip 0.17cm} c @{\hskip 0.17cm} c} 
 \hline
 Code - Training Set & $z$ &$\zeta$ & $R_\text{mfp}$ & $\log T_\text{vir}$\\ [0.5ex] 
 \hline\hline
 21CMMC & 9 & $41.28^{+24.85}_{-13.43}$ & $13.38^{+4.28}_{-5.15}$ & $4.59^{+0.37}_{-0.32}$ \\ 
 \hline
 ANN - 100 LHS & 9 & $45.47^{+25.19}_{-17.18}$ & $12.13^{+5.71}_{-5.05}$ & $4.54^{+0.47}_{-0.28}$ \\
 \hline
 ANN - 1000 LHS & 9 & $42.52^{+26.18}_{-13.74}$ & $12.89^{+4.63}_{-5.29}$ & $4.57^{+0.40}_{-0.31}$ \\
 \hline
 ANN - 10000 LHS & 9 & $42.21^{+25.42}_{-14.12}$ & $13.18^{+4.46}_{-5.14}$ & $4.58^{+0.39}_{-0.31}$\\
 \hline\hline
 21CMMC & 8 & $39.64^{+31.90}_{-16.11}$ & $14.99^{+2.98}_{-3.64}$ & $4.61^{+0.21}_{-0.23}$\\ 
 \hline
 ANN - 100 LHS & 8 & $43.06^{+26.16}_{-17.38}$ & $14.58^{+3.47}_{-3.90}$ & $4.64^{+0.19}_{-0.25}$\\
 \hline
 ANN - 1000 LHS & 8 & $42.71^{+31.30}_{-18.67}$ & $14.67^{+3.19}_{-4.26}$ & $4.62^{+0.21}_{-0.23}$\\
 \hline
 ANN - 10000 LHS & 8 & $39.78^{+31.68}_{-16.22}$ & $14.61^{+3.15}_{-4.05}$ & $4.60^{+0.22}_{-0.23}$\\ 
 \hline\hline
 21CMMC & 8,9,10 & $31.08^{+8.70}_{-6.04}$ & $15.15^{+2.86}_{-3.21}$ & $4.51^{+0.17}_{-0.17}$ \\ \hline
 ANN - 100 LHS & 8,9,10 & $31.51^{+8.57}_{-6.32}$ & $15.86^{+2.47}_{-3.62}$ & $4.49^{+0.16}_{-0.19}$ \\ \hline
 ANN - 1000 LHS & 8,9,10 & $31.18^{+8.47}_{-6.08}$ & $14.97^{+2.91}_{-3.78}$ & $4.51^{+0.16}_{-0.17}$ \\ \hline
 ANN - 64 gridded & 8,9,10 & $32.46^{+13.90}_{-5.72}$ & $12.52^{+3.47}_{-6.13}$ & $4.61^{+0.11}_{-0.13}$ \\ \hline
 ANN - 125 gridded & 8,9,10 & $30.17^{+6.78}_{-5.04}$ & $12.97^{+4.09}_{-3.69}$ & $4.50^{+0.15}_{-0.16}$ \\ \hline
 ANN - 1000 gridded & 8,9,10 & $31.32^{+7.52}_{-5.20}$ & $13.94^{+3.80}_{-4.68}$ & $4.50^{+0.16}_{-0.16}$ \\ \hline
\end{tabular}
\label{table: statistics summary}
\end{center}

\end{table}

\begin{figure}
\includegraphics[width=0.5\textwidth]{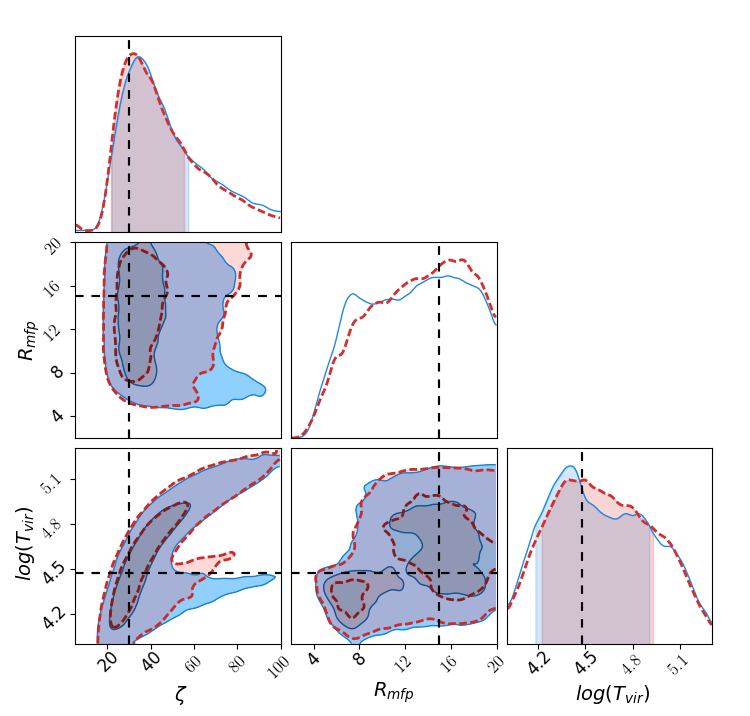}
\caption{Comparison between the recovered $1\sigma$ and $2\sigma$ confidence regions of 21CMMC (red dashed lines) and the ANN emulator (blue solid lines) at $z = 9$. The ANN uses 1000 LHS for the training set and a $10^4$ training iterations. The dotted lines indicate the true parameter values $(\zeta, R_\text{mfp}, \log T_\text{vir}) = (30, 15, 4.48)$.}
\label{fig: mcmc 1000LHS z9}
\end{figure}

\begin{figure}
\includegraphics[width=0.5\textwidth]{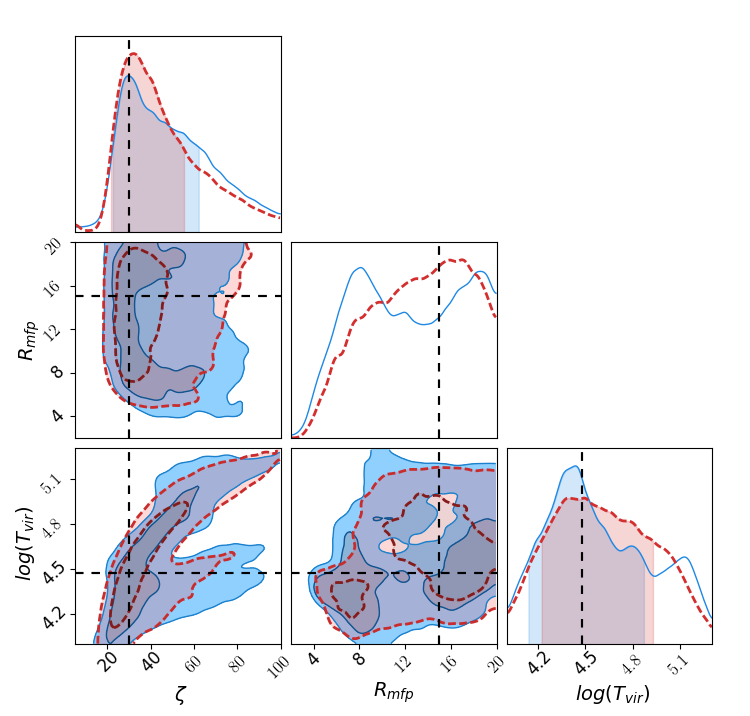}
\caption{Comparison between the recovered $1\sigma$ and $2\sigma$ confidence regions of 21CMMC (red dashed lines) and the ANN emulator (blue solid lines) at $z = 9$. The ANN uses 100 LHS for the training set and a $10^4$ training iterations. The dotted lines indicate the true parameter values $(\zeta, R_\text{mfp}, \log T_\text{vir}) = (30, 15, 4.48)$.}
\label{fig: mcmc 100LHS z9}
\end{figure}

\begin{figure}
\includegraphics[width=0.5\textwidth]{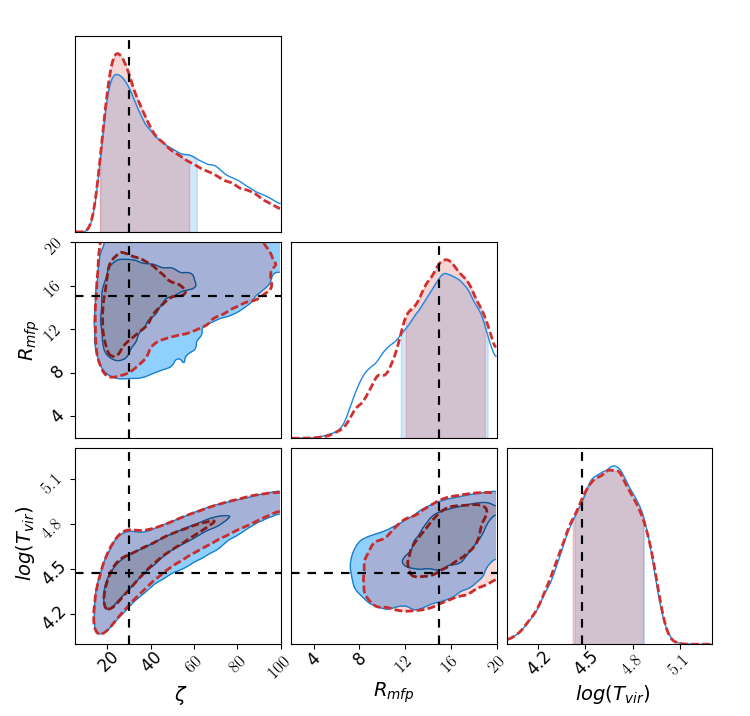}
\caption{Comparison between the recovered $1\sigma$ and $2\sigma$ confidence regions of 21CMMC (red dashed lines) and the ANN emulator (blue solid lines) at $z = 8$. The ANN uses 1000 LHS for the training set and a $10^4$ training iterations. The dotted lines indicate the true parameter values $(\zeta, R_\text{mfp}, \log T_\text{vir}) = (30, 15, 4.48)$.}
\label{fig: mcmc 1000LHS z8}
\end{figure}

\begin{figure}
\includegraphics[width=0.5\textwidth]{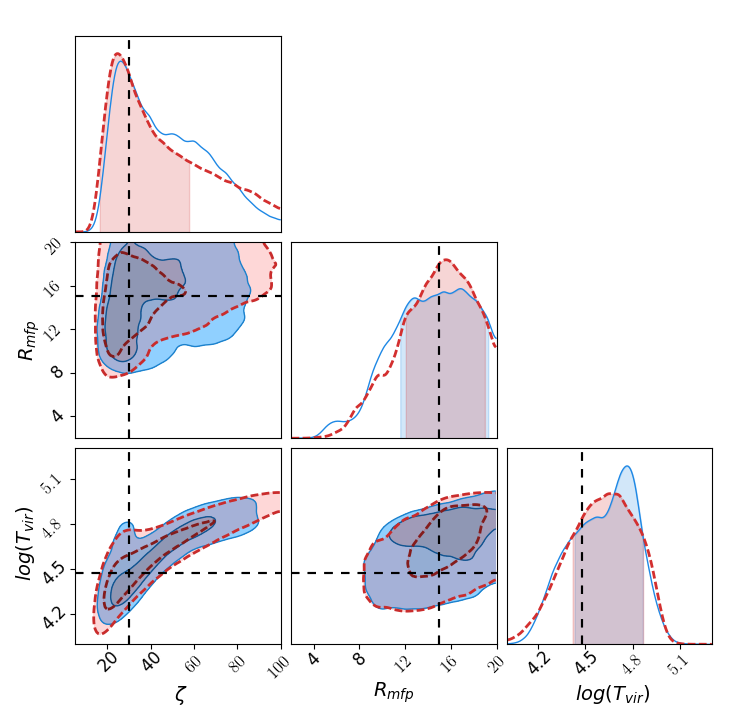}
\caption{Comparison between the recovered $1\sigma$ and $2\sigma$ confidence regions of 21CMMC (red dashed lines) and the ANN emulator (blue solid lines) at $z = 8$. The ANN uses 100 LHS for the training set and a $10^4$ training iterations. The dotted lines indicate the true parameter values $(\zeta, R_\text{mfp}, \log T_\text{vir}) = (30, 15, 4.48)$.}
\label{fig: mcmc 100LHS z8}
\end{figure}

Figures \ref{fig: mcmc 1000LHS z9} to \ref{fig: mcmc 100LHS z8} show the comparison between the brute-force parameter estimation as the red dashed contours and our ANN emulation using a variety of training set sizes at redshift $z = 9$ and $z = 8$ as the solid blue contours.
For both redshifts, we show the one and two sigma contours obtained for 100 and 1000 LH samples as well as the marginalized posteriors convolved with a gaussian smoothing kernel.
As our posterior 1D marginalized parameter distributions are not found to be gaussian, we compute the median and the 68\% confidence interval defined by the region between the 16th and 84th percentile as our summary statistics in Table \ref{table: statistics summary}. 
We find excellent agreement between our method and 21CMMC for training sets of $10^3$ and $10^4$ samples at both redshifts, and good agreement with 100 samples.

We observe that errors retrieved by our network can be smaller than those obtained by 21CMMC, this is due to systematics.
During the training period, our ANN constructs a model which approximates the 21cmFAST model and we proceed to sample the likelihood of the approximation.
Therefore, assuming convergence of the chains, any difference between the recovered 68\% confidence intervals are most likely due to systematic difference between the two models that are sampled.
We estimate that we are subject to these systematic effects on the 1\% - 10\% level for large to small training sets, as per Figures \ref{fig: training iter vs loss} and \ref{fig: comparison of errors}.

The $\zeta - \log T_\text{vir}$ panels in Figures \ref{fig: mcmc 1000LHS z9} and \ref{fig: mcmc 100LHS z9} show that the neural network is sensitive to the same multi-modality found by 21CMMC, which is illustrated by the stripe feature at low $T_\text{vir}$ and high $\zeta$.
This region represents a less massive galaxies with a brighter stellar population, which can mimic our fiducial observation. 
Such a galaxy population would ionize the IGM earlier and thus by combining  multiple redshifts and adding information about the evolution of the ionization process, this degeneracy ought to be lifted.
Similarly, the $R_\text{mfp} - \log T_\text{vir}$ panel shows a clear bimodal feature for both 21CMMC and our neural network.
Comparing to the results at $z = 8$ in Figures \ref{fig: mcmc 1000LHS z8} and \ref{fig: mcmc 100LHS z8}, we see this multi-modal behaviour disappearing, which suggest that this degeneracy can be lifted by adding information in multiple redshift bins.
Despite a clear downgrade of the fit to the brute-force method in the shape of both the 2D contours and the 1D marginalized posteriors, the training set using 100 samples still encloses the true parameter values of the observation in the 68\% confidence interval as indicated in Table \ref{table: statistics summary}.


\subsection{Multiple Redshift Parameter Constraints}

\begin{figure}
\includegraphics[width=0.5\textwidth]{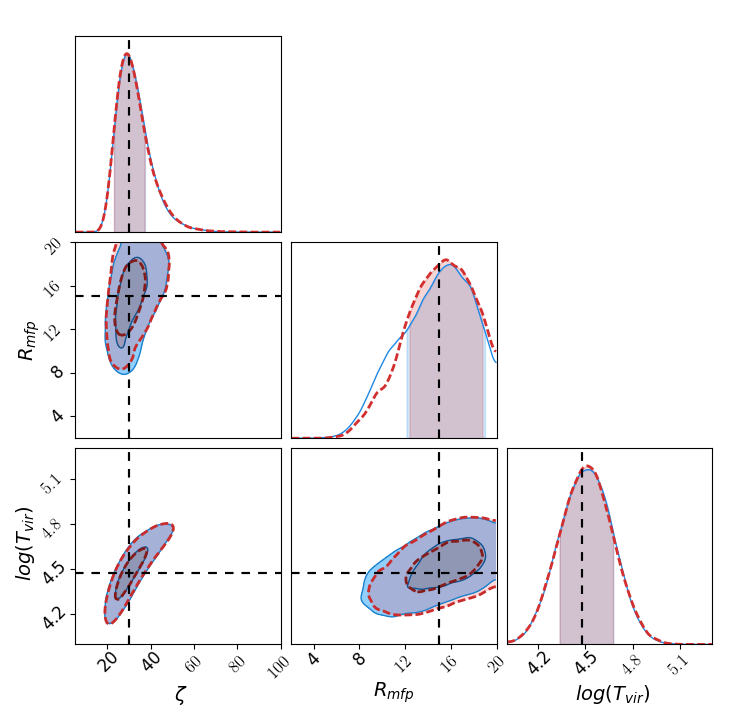}
\caption{Comparison between the recovered $1\sigma$ and $2\sigma$ confidence regions of 21CMMC (red dashed lines) and the ANN emulator (blue solid lines) combining redshifts $z = 8$, $z = 9$, and $z = 10$. The ANN uses 1000 LHS for the training set at each redshift and a $10^4$ training iterations. The dotted lines indicate the true parameter values $(\zeta, R_\text{mfp}, \log T_\text{vir}) = (30, 15, 4.48)$.}
\label{fig: mcmc 1000LHS multi}
\end{figure}

\begin{figure}
\includegraphics[width=0.5\textwidth]{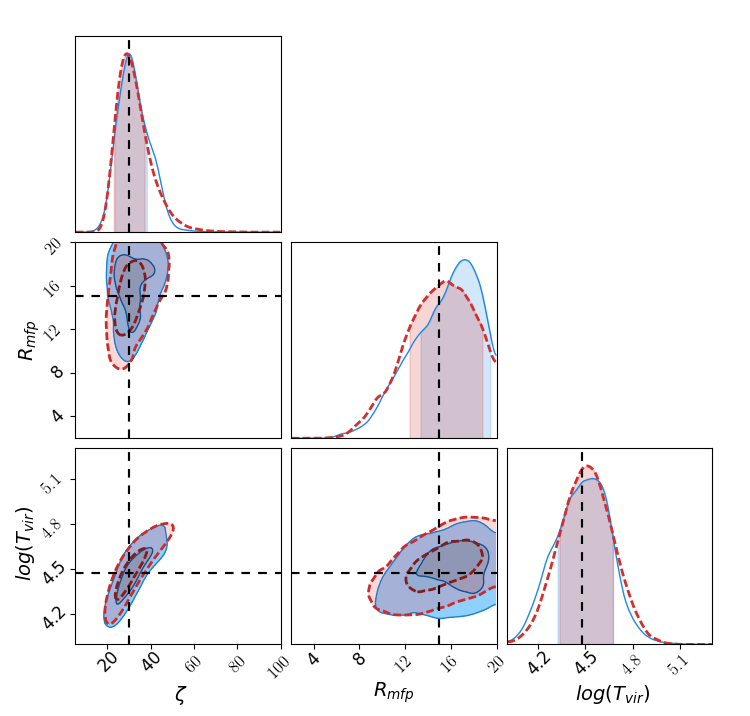}
\caption{Comparison between the recovered $1\sigma$ and $2\sigma$ confidence regions of 21CMMC (red dashed lines) and the ANN emulator (blue solid lines) combining redshifts $z = 8$, $z = 9$, and $z = 10$. The ANN uses 100 LHS for the training set at each redshift and a $10^4$ training iterations. The dotted lines indicate the true parameter values $(\zeta, R_\text{mfp}, \log T_\text{vir}) = (30, 15, 4.48)$.}
\label{fig: mcmc 100LHS multi}
\end{figure}

Figures \ref{fig: mcmc 1000LHS multi} and \ref{fig: mcmc 100LHS multi} show the contraints obtained when combining observations in three redshift bins at $z = 8$, 9 and 10 for training sets of 1000 and 100 samples per redshift respectively.
As noted in the previous section, adding information about the evolution of the reionization process lifts some of the degeneracies in our recovered parameter constraints and both multi-modal features in the $\zeta - \log T_\text{vir}$ and the  $R_\text{mfp} - \log T_\text{vir}$ panels could be lifted.
Of note is that combining multiple redshift bins highly improves the fit of the neural network trained on only 100 samples per redshift.
We find that all our fiducial parameter values are well within the 68\% confidence interval set our by the median and it's 16th and 84th percentile for even this sparse training set.

Additionally, we compare the inference of a network trained on gridded training sets with similar sizes to our LH sampled training sets. 
Both $5^3$ and $10^3$ training sets recover similar constrains as the 100 LHS and 1000 LHS training sets, consistent with our findings in Section \ref{sec: power}.
However, we observe a clear deterioration of the predictive power as we reduce the number of gridded training parameters to 4 points per parameter. 
Although the fiducial parameter values are recovered within the 16th to 84th percentile in Table \ref{table: statistics summary}, we fail to recover the fiducial values within the $2\sigma$ contours for $4^3$ points.

\subsection{Applications}

With a speed-up of $\sim 3$ orders of magnitude, 21cm power spectrum emulation can be used for a variety of new or existing analyses, and we aim here to highlight some potential uses:

(\textit{i}) 21cm experimental design studies (eg. \cite{greig2015a}) use much the same principle as our model parameter inference outlined above. 
By varying the experimental layout or survey strategy, we effectively vary the noise power spectrum $P_N(k)$ in Equation \ref{eq:loglikelihood}, and can thus fit the optimal layout or survey strategy. 
These studies require fast model evaluations in order to be able to compare a multitude of survey strategies and experimental design. 

(\textit{ii}) We find that using small training sets of 100 model evaluations, our emulation recovers parameter constraints to a similar degree of accuracy as those obtained when evaluating the model at each point in the chain.
This may open up the possibility to move away from semi-numerical models such as \texttt{21cmFast} and for the first time use radiative transfer codes \citep{Ciardi2003,Iliev2006,baek2009,baek2010} in EoR parameter searches.
\citet{Semelin2017} have recently produced a first database of 45 evaluations of their radiative transfer code to provide 21cm brightness temperature lightcones evaluated on a 3D grid. 
The power spectra extracted from this database could be used as a training set for an ANN emulator. 
However, our analysis suggests that training sets with lower than 100 samples should be used with caution.

(\textit{iii}) In addition to determining the best fit parameters of any given model, we would like to quantify the degree of belief in our model in the first place. 
Future data will be abundant, and as such we would like to be able to use it to inform us about the choice of model that best fits the data. 
Here too, the computational speed that emulation provides can be of use.
Bayesian model comparison requires the computation of the evidence as the integral of the likelihood times the prior over all of parameter space.
Nested sampling algorithms such as \texttt{MultiNest} \citep{Feroz2009} provide an estimate for the evidence of a particular model together with the evaluation of the posterior, and thus benefits greatly from fast power spectrum computations. 

(\textit{iv}) The output nodes of the neural network treats each k-bin of the 21cm power spectrum separately. 
The weights of the trained network thus act to correlate the values in each k-bin according to the training set.
There is therefore no restriction to predict other observables that are correlated to the 21cm power spectrum using the same emulator.
The same network could thus encode the skewness or bispectrum of the 21cm fluctuations at the same time assuming the inclusion of these functions in the training sets.

\section{Conclusion}
\label{sec: conclusion}
With the advent of next generation telescopes such as MWA, HERA and the SKA, a first detection of the cosmic 21cm signal from the Epoch of Reionization is expected to be made within the next few years.
In order to infer EoR parameters from these observations, expensive model evaluations are needed to compare to the data.
One avenue to reduce the computational cost of model evaluations is by using machine learning techniques to emulate the model.
We show that emulating the models using artificial neural networks can speed up the model evaluations significantly, while maintaining a high degree of accuracy.
We use our ANN to train on a series of training sets which consist of 21cm power spectrum evaluations produced by the semi-numerical code \texttt{21cmFast}.
As the limiting factor now becomes the creation of the training set, we study the evolution of the error on the power spectrum predictions as a function of the training size and find that as few as 100 model evaluations may be sufficient to recover reasonable constraints on the parameters, especially when combining information across multiple redshift bins.


\section*{Acknowledgments}

We would like to thank Benoit Semelin, William Jennings, Alan Heavens, and Conor O'Riordan for their helpful conversations and suggestions.
We acknowledge use of Scikit-learn \citep{scikit-learn}.
JRP is pleased to acknowledge support under FP7-PEOPLE-2012-CIG grant number 321933-21ALPHA, and from the European Research Council under ERC grant number 638743-FIRSTDAWN.
CJS acknowledges the National Research
Fund, Luxembourg grant \say{Cosmic Shear from the Cosmic Dawn - 8837790}.

\bibliography{NNbib}


\label{lastpage}

\end{document}